\documentclass[twocolumn]{autart}    
\usepackage{amsmath}
\usepackage{color}
\usepackage{graphicx}
\usepackage{amsfonts}
\usepackage{amssymb}
\usepackage{hyperref}
\usepackage{color}
\usepackage{graphicx,subfigure}

 \newtheorem{theorem}{Theorem}
\newtheorem{definition}{Definition}
\newtheorem{proposition}{Proposition}
\newtheorem{lemma}{Lemma}
\newtheorem{corollary}{Corollary}
\newtheorem{example}{Example}

\newcommand{\rank} {\mbox{\rm rank}\,}
\newcommand{\diag} {\mbox{\rm diag}\,}
\newcommand{\grk} {\mbox{\rm grk}\,}

\begin{document}

\begin{frontmatter}

\title{Zeros of Networked Systems with Time-invariant  Interconnections} 

\thanks[footnoteinfo]{This paper was not presented at any IFAC
meeting. Corresponding author is Mohsen Zamani.}

\author[label1]{Mohsen Zamani,}
\author[label2]{Uwe Helmke,}
\author[label3]{Brian D. O. Anderson}
\address[label1]{Research School of Engineering, Australian National University, Canberra, ACT 0200, Australia. (e-mail: mohsen.zamani@anu.edu.au)}
\address[label2]{Institute of Mathematics, University of W\"{u}rzburg, 97074 W\"{u}rzburg, Germany (email: helmke@mathematik.uni-wuerzburg.de)}
\address[label3]{ Research School of Engineering, Australian National University, Canberra, ACT 0200, Australia and Canberra Research Laboratory, National ICT Australia Ltd., PO Box 8001, Canberra, ACT 2601, Australia. (e-mail:
brian.anderson@anu.edu.au) }


\begin{abstract}
This paper studies zeros of networked linear systems with time-invariant interconnection  topology. While  the  characterization of zeros is given for both heterogeneous and homogeneous networks,   homogeneous networks are explored in greater detail. In the current paper, for homogeneous networks with time-invariant interconnection dynamics,  it is   illustrated how the zeros of each individual agent's system description
and zeros definable from the interconnection dynamics contribute to generating zeros of the whole network.  We also demonstrate how zeros of  networked systems and those of their  associated blocked versions are related.
\end{abstract}

\end{frontmatter}

\section{Introduction}\label{intro}
Recent developments of enabling technologies such as communication systems, cheap
computation equipment and sensor platforms have given great impetus to the creation
of  networked systems. Thus, this area has attracted significant
attention worldwide and researchers have studied networked systems from
different perspectives (see e.g. \cite{sinopoli2003distributed},
\cite{olfati2002distributed}, \cite{tanner2003stable}). In particular,
in view of the recent chain of events \cite{gorman2009electricity},
\cite{falliere2011w32} and \cite{rid2012cyber}, the issues of security and cyber
threats to the networked systems have gained  growing attention. This paper uses  system theoretic approaches  to deal with problems involved with the security of networks.

Recent works have shown that control theory can be used as an effective tool to detect and mitigate the effects of cyber  attacks on the networked systems;  see for example  \cite{sinapoli2012}, \cite{cardenas2011attacks}, \cite{gupta2010optimal}, \cite{amin:09}, \cite{govinndarasu}, \cite {Teixeira202} and the references listed therein. The authors of   \cite {Teixeira202} have introduced the concept of \textit{ zero-dynamics attacks}
and  shown how attackers can use knowledge of networks' zeros
to produce control commands such that they are not detected as security threats. Thus,  zeros of networks provide valuable
information relevant to detecting cyber attacks. Though  various aspects of the dynamics of networked systems have been extensively
studied in the literature \cite{Ren2007,Olfati2007,fax2004}, to the authors' best knowledge
the zeros of networked systems have not been studied in any detail \cite{zamanihlemke2013}.

This paper  examines  the zeros of networked systems in more depth. Our focus is on networks of finite-dimensional linear
dynamical systems that arise through static interconnections of a finite number of
such systems. Such models arise naturally in  applications
of linear  networked systems, e.g. for cyclic pursuit \cite{MarshallBrouckeFrancis};
 shortening flows in image processing \cite{Bruckstein}, or for the
 discretization of partial differential equations
 \cite{brockett-willems}.

Our ultimate goal is to analyze the zeros of networked systems with
periodic, or more  generally time-varying interconnection topology.
An important tool for this analysis is  blocking or lifting technique
for networks with time-invariant interconnections.  Note that blocking of linear time-invariant systems is  useful  in design of controllers for linear periodic systems  as shown by
\cite{chenB95} and \cite{Khargoneckar85}. 
References \cite{Bolzern86}, \cite{Grasselli88},
  \cite{zamani2011} and \cite{chen2010} have analyzed   zeros
    of blocked systems obtained from blocking of
  time-invariant  systems. Their works were  extended by
  \cite{zamani2011}, \cite{chen2010}.
  However, these earlier contributions  do not take  any underlying network structure
  into consideration. In this paper,  we introduce some results that
  provide a first step in that direction.

The structure of this paper is as follows. First, in Section
\ref{sec:models} we introduce state-space and higher order
polynomial system models for time-invariant networks of linear
systems. A central result used is the strict system equivalence
between these different system representations.  Moreover, we
completely characterize  both finite and infinite zeros of arbitrary
heterogeneous networks. For homogeneous networks of identical SISO
systems more explicit results are provided in Section \ref{sec:homog}. Homogeneous networks with a
circulant coupling topology are studied as well. In Section  \ref{sec:blocking},
a relation between the transfer function of the blocked system and
the transfer function of the associated unblocked system is
explained.  We then relate the zeros of  blocked networked systems
to those  of the  corresponding unblocked systems, generalizing work by
\cite{zamani2011}, \cite{chen2010},\cite{MOHSEN-SCLpaper}.  Finally, Section \ref{sec:conclusion} provides the concluding remarks.

\section{Problem Statement and Preliminaries}
\label{sec:models}
We consider networks of $N$ linear systems, coupled
through constant interconnection parameters. Each agent is assumed to
have the state-space representation as a linear discrete-time  system
\begin{equation}
\begin{split}
\label{sys1}
x_i(t+1) &=A_ix_i(t)+B_iv_i(t)\\
w_i(t)   &=C_ix_i(t),\; i=1,\dots,N.
\end{split}
\end{equation}
Here,  $A_i\in \mathbb{R}^{n_i\times n_i}$, $B_i\in
\mathbb{R}^{n_i\times m_i}$ and $C_i\in \mathbb{R}^{p_i\times
  n_i}$ are the associated system matrices. We assume that each system is reachable and
observable and that the agents are interconnected by static
coupling laws
\begin{equation*}
v_i(t)=\sum_{j=1}^NL_{ij}w_j(t)+R_iu(t)\in \mathbb{R}^{m_i}
\end{equation*}
with $L_{ij}\in \mathbb{R}^{m_i\times p_j}$, $R_i\in
\mathbb{R}^{m_i\times m}$  and $u(t)\in \mathbb{R}^m$ denoting an external input
applied  to the whole
network. Further, we assume that there is a $p$-dimensional interconnected output  given by
\begin{equation*}
y(t) = \sum_{i=1}^NS_iw_i(t) +Du(t)
 \;\mbox{ with }\;
S_i\in \mathbb{R}^{p\times p_i}, \; i=1,\dots,N.
\end{equation*}
Define $\overline{m}=\sum_{i=1}^{N}m_i$,
$\overline{p}=\sum_{i=1}^{N}p_i$, $\overline{n}=\sum_{i=1}^{N}n_i$
and \textit{coupling matrices}
\begin{eqnarray*}
L&=& (L_{ij})_{ij}\in \mathbb{R}^{\overline{m}\times\overline{p}} \quad
R=\begin{pmatrix}R_1\\ \vdots\\ R_N\end{pmatrix}\in
\mathbb{R}^{\overline{n}\times m}\\
S&=&(S_1,\dots,S_N)\in \mathbb{R}^{p\times \overline{p}}
\quad
D\in \mathbb{R}^{p\times m}
\end{eqnarray*}
as well as \textit{node matrices}
\begin{equation}\label{eq:nodematrices}
\begin{split}
A  &=\diag
(A_1, \ldots, A_N),
\quad
B  =\diag
(B_1, \ldots, B_N)\\
C  &=\diag
(C_1, \ldots, C_N),\quad
x(t) :=
 \begin{pmatrix}
 x_1(t)\\\vdots\\x_N(t)
 \end{pmatrix}\in \mathbb{R}^{\overline{n}}.
\end{split}
\end{equation}

Then the closed-loop system is
\begin{equation}\label{eq:system}
\begin{split}
x(t+1) &= {\mathbf{A}}x(t)+ {\mathbf{B}}u(t)\\
y(t) &= {\mathbf{C}}x(t)+Du(t),
\end{split}
\end{equation}
with matrices
\begin{equation}
\mathbf{A} := A + B L C
\quad
\mathbf{B} := B R,
\quad
{\mathbf{C}} := SC.
\end{equation}

One can also start by assuming that each  system (\ref{sys1}) is defined in terms of Rosenbrock-type equations \cite{Rosenbrock1970} i.e. by systems of higher order
difference equations
\begin{equation}
\label{SIC24}
\begin{array}{rcl}
T_i(\sigma)\xi_i(t) &=& U_i(\sigma) v_i(t)\\
w_i(t) &=& V_i(\sigma) \xi_i(t).
\end{array}
\end{equation}

Here $\sigma$ denotes the forward shift operator that acts on
sequences of vectors $(\xi(t))_{t}$ as $(\sigma \xi(t))
=\xi(t+1)$. Furthermore, $T_i,U_i,V_i$ denote polynomial matrices of
sizes $T_i(z) \in \mathbb{R}[z]^{r_i\times r_i}, U_i(z) \in
\mathbb{R}[z]^{r_i\times m_i}$ and $V_i(z) \in \mathbb{R}[z]^{p_i\times
  r_i}$, respectively. We always assume  that $T_i(z)$ is  \textit{nonsingular},
i.e. that $\det T_i(z)$ is not the zero polynomial. Moreover, the
system (\ref{SIC24}) is assumed to be strictly proper, i.e. we assume
that the associated transfer function
\begin{equation}
\begin{split}
\label{SIC20}
G_i(z) &= V_i(z)T_i(z)^{-1}U_i(z)
\end{split}
\end{equation}
is strictly proper.  Following Fuhrmann \cite{Fuhrmann1977}, any
strictly proper system of higher order
difference equations has an associated state-space realization
$(A,B,C)$, the so-called \textbf{shift realization}, such that
the polynomial matrices
\begin{equation}
  \label{eq:systemmatrix}
 \begin{pmatrix}
zI-A &-B\\
C& 0\\
\end{pmatrix},\quad  \begin{pmatrix}
T(z)&-U(z)\\
V(z)& 0\\
\end{pmatrix}
\end{equation}
 are strict system equivalent \cite{Fuhrmann1977}. If the  first order representation (\ref{sys1}) is strict
system equivalent to the higher order system (\ref{SIC24}) then of
course the associated transfer functions coincide, i.e. we have
\begin{equation}
\begin{split}
\label{SIC20}
C_i(zI-A_i)^{-1}B_i &= V_i(z)T_i(z)^{-1}U_i(z).
\end{split}
\end{equation}

Throughout this paper we assume that the first order and higher
order representations  i.e. the  systems (\ref{sys1}) and (\ref{SIC24}),  are chosen to be of minimal
order, respectively. This is equivalent to the controllability and
observability of the shift realizations (\ref{sys1}) associated with
these representations (\ref{SIC24}). It is also equivalent to the  simultaneous left coprimeness
of $T_i(z),U_i(z)$ and the right coprimeness of $T_i(z), V_i(z)$.
Proceeding as above, define polynomial matrices
\begin{equation}
\label{PMD25}
\begin{array}{rcl}
T(z) &=& \diag  (T_1(z), \ldots, T_N(z) ) \in \mathbb{R}[z]^{\overline{r} \times \overline{r}}\\
\end{array}
\end{equation}
and similarly for $V(z)$ and $U(z)$. Here $\overline{r}=\sum_{i=1}^{N}r_i$.
Using this  notation,    we  write all $N$ systems of  (\ref{SIC24})  in the  matrix form as
\begin{equation}
\label{PMD28}
\left(\begin{array}{c}
0\\
I\\
\end{array}\right)w(t) = \left(\begin{array}{cc}
T(\sigma) & -U(\sigma)\\
V(\sigma) & 0\\
\end{array}\right)\left(\begin{array}{c}
\xi(t)\\
v(t)\\
\end{array}\right),
\end{equation}
where $w(t)=\left(
              \begin{array}{cccc}
                w_1(t)^{\top} & w_2(t)^{\top} & \ldots & w_N(t)^{\top} \\
              \end{array}
            \right)^{\top}
 $ and similarly for $\xi(t)$ and $v(t)$.
Then we have the left- and right coprime factorizations of the
 $\overline p \times \overline m$ {\it
  node transfer function} as
 \begin{equation*}
G(z)=C (zI-A)^{-1}B=V(z)T(z)^{-1}U(z).
\end{equation*}
The interconnections are given, as before, by
\begin{equation*}
\begin{array}{rcl}
v(t) &=& Lw(t) + Ru(t) \\
y(t) &=& S w(t)+Du(t).\\
\end{array}
\end{equation*}

The  resulting network representation then becomes
 \begin{equation}
\label{eq:polysystem}
\left(\begin{array}{c}
0\\
I\\
\end{array}\right)y(t) = \left(\begin{array}{cc}
T(\sigma)-U(\sigma)LV(\sigma) & -U(\sigma)R\\
SV(\sigma) & D\\
\end{array}\right)\left(\begin{array}{c}
\xi(t)\\
u(t)\\
\end{array}\right)
\end{equation}
with the $p\times m$  {\it network transfer function} defined as

\begin{equation}\label{PMD31}
\begin{split}
\Gamma(z)&= \mathbf{C}(zI-\mathbf{A})^{-1}\mathbf{B}+D\\
&= SV(z)(T(z)-U(z)LV(z))^{-1}U(z)R+D.
\end{split}
\end{equation}

The connection between the state-space and the  polynomial matrix
representations (\ref{eq:system}) and (\ref{eq:polysystem}), respectively,
is clarified by the following result. This theorem implies that important
system-theoretic properties such as reachability and observability, as
well as the poles and zeros of the networked system (\ref{eq:system}) can all  be  characterized by the polynomial
system matrix (\ref{eq:polysystem}).\\

\begin{theorem}[\cite{fuhe2013}]\label{sysequiv}
The interconnected systems (\ref{eq:system}) and (\ref{eq:polysystem})
are strict system equivalent. In particular, for each $q\geq \max
(\overline{n}, \overline{r})$ there exist unimodular polynomial matrices
$E(z),F(z)$ such that
\begin{equation*}
\begin{split}
&E(z)\begin{pmatrix}
I_{q-\overline{n}}&0&0\\
0&zI-\mathbf{A}&-\mathbf{B}\\
0&\mathbf{C}& D\\
\end{pmatrix} F(z)=\\&\begin{pmatrix}
I_{q-\overline{r}}&0&0\\
0&T(z)-U(z)LV(z) &-U(z)R\\
0&SV(z)& D\\
\end{pmatrix}.
\end{split}
\end{equation*}
\end{theorem}

As a consequence of Theorem \ref{sysequiv} we derive a complete characterization for the zeros of
the system (\ref{eq:system}). We first present an extension
of the classical
definition of the zeros \cite{kailath} to the higher
order system \eqref{SIC24}. Note that the \textbf{normal rank} $\grk \:G(z)$ of a rational matrix
function $G(z)$ is defined as
\begin{equation*}
\grk\;G(z)=\max \{\rank \;G(z)\;|\;z\in \mathbb{C}, G(z)\neq
\infty\}.
\end{equation*}

\begin{definition}\label{def:def1}
Let $U(z),V(z),T(z)$ be polynomial matrices with
$T(z)\in\mathbb{R}[z]^{\overline r \times \overline r}$
nonsingular such that the $\overline p \times \overline m$ node transfer function $V(z)T(z)^{-1}U(z)$ is
strictly proper; let $D$ be a constant matrix. A \textbf{finite zero} of the  polynomial system matrix
\begin{equation}
  \label{eq:systemmatrix}
\Pi(z)=\begin{pmatrix}
T(z) &-U(z)\\
V(z)& D\\
\end{pmatrix}
\end{equation}
is any complex number
$z_0\in \mathbb{C}$ such that
\begin{equation*}
\rank\; \Pi(z_0) < \grk\;  \Pi(z)
\end{equation*}
holds.   $ \Pi(z)$ is said to have a  \textbf{zero at infinity} if
\begin{equation*}
\overline r+\rank\;  D < \grk\; \Pi(z).
\end{equation*}
\end{definition}

As a consequence of Fuhrmann's result \cite{Fuhrmann1977}, a
 polynomial system matrix (\ref{eq:systemmatrix}) has a finite or
 infinite zero if and only if the polynomial matrix

\begin{equation*}
\Sigma(z)=\begin{pmatrix}
zI-A &-B\\
C& D\\
\end{pmatrix}
\end{equation*}
of the associated shift realization
$(A,B,C, D)$ has a finite or infinite
 zero. Theorem \ref{sysequiv} thus leads to a complete
 characterization
 of the zeros for the interconnected system \eqref{eq:system}  as stated in the subsequent theorem.  We emphasize that the characterization of the zeros in the
 subsequent Theorem \ref{MAINA}
holds for any interconnection matrices and does not require any assumptions on reachability or observability of
the network, except of those for the individual  node systems.\\

\begin{theorem}[\cite{fuhe2013}]
\label{MAINA}
Consider the strictly proper   node transfer function $G(z)$ with
minimal representations (\ref{eq:nodematrices}) as
\begin{equation*}
\label{SIC36A}
G(z) = C (zI-A)^{-1}B = V(z)T(z)^{-1}U(z).
\end{equation*}
Let $L,R,S,D$ be any arbitrary constant interconnection matrices of the proper dimensions and
\begin{equation*}
\Gamma(z)=SV(z)(T(z)-U(z)LV(z))^{-1}U(z)R+D
\end{equation*}
denote the network transfer function.
Assume that $G(z)$ is represented by  a polynomial left coprime matrix fraction
description (MFD) as

$$G(z)=D^{-1}_L(z)N_L(z).$$

 Then
\begin{enumerate}
\item For all $z\in \mathbb{C}$
\begin{equation*}
\begin{split}
& \rank \left(\begin{array}{cc}
zI-\mathbf A  & -\mathbf B \\
\mathbf C & D
\end{array}\right)\\
&=\overline{n}-\overline{r}+\rank \left(\begin{array}{cc}
T(z) -U(z)LV(z) & -U(z)R\\
SV(z) & D \\
\end{array}\right).\end{split}
\end{equation*}

 \item For all $z \in \mathbb C$
 \begin{equation*}
 \begin{split}
 & \rank \left(\begin{array}{cc}
 zI-A-B L C  & -B R\\
 SC & D
 \end{array}\right)\\
 &=\overline{n}-\overline{p}+\rank \left(\begin{array}{cc}
 D_L(z) -N_L(z)L & -N_L(z)R\\
 S& D \\
 \end{array}\right).\end{split}
 \end{equation*}
\item
$(\mathbf A, \mathbf B,
\mathbf C,D)$ has a finite zero at $z_0\in \mathbb{C}$ if and only if
\begin{equation*}
\begin{split}
&\rank \left(\begin{array}{cc}
T(z_0) -U(z_0)LV(z_0) & -U(z_0)R\\
SV(z_0) & D \\
\end{array}\right)<\\&\overline{r}+\grk\;\Gamma (z).
\end{split}
\end{equation*}


\item $(A+B LC, B R, \label{partx}
SC,D)$ has a zero at infinity if and only if
\begin{equation*}
\rank\; D < \grk\; \Gamma (z).
\end{equation*}
In particular, if $D$ has full-row rank or full-column rank, then $(A+B LC, B R,
SC,D)$ has no infinite zero.
 \end{enumerate}
\end{theorem}

\section{Zeros of Homogeneous Networks}\label{sec:homog}

The preceding result has a nice simplification in the case of
\textbf{homogeneous networks  of SISO agents}, i.e. where the node systems $(A_i,B_i,C_i)$ are single input single output systems with identical transfer function.  Let us define the  {\it interconnection transfer function} as
 \begin{equation*}
\phi(z)=S(zI-L)^{-1}R+D.
\end{equation*}

The next theorem relates the zeros of the system \eqref{eq:system} to those of the interconnection dynamics \footnote{The term interconnection dynamics
is partly a misnomer. There is no dynamics separate to that included within the agent description, and the interconnecting matrices are all constant. The transfer function $\phi(z)$ is a theoretical construct: it is the transfer function from $u(t)$ to $y(t)$ resulting when every system is replaced by $z^{-1}$.} defined by the quadruple $(L,R,S,D)$. Before we provide this main result, we need to state the following lemma regarding the generic rank of $\Gamma (z)$.

\begin{lemma}\label{lem:normalrank}
Assume that  $(A_i,b_i,c_i)$ are scalar SISO systems with identical transfer function $g(z)=c_i(zI-A_i)^{-1}b_i$. Let $L,R,S,D$ denote any constant interconnection matrices of the proper dimensions  and $\phi(z)=S(zI-L)^{-1}R+D$ be the interconnection transfer function. Then the following equality holds.
$$
\grk\; \Gamma(z)=\grk\; \phi(z).
$$
\end{lemma}
\noindent \textbf{Proof.}
Consider any coprime factorization $g(z)=\frac{p(z)}{q(z)}$ of the
strictly proper transfer function $g(z)$, having McMillan degree
$n$. Define $h(z)=g(z)^{-1}=\frac{q(z)}{p(z)}$.  We know that
\begin{equation}\label{eq:lem1eq1}
\begin{split}
\grk\; \Gamma(z)&=\grk\left(\begin{array}{cc}
zI-\mathbf A & -\mathbf B \\
\mathbf C & D
\end{array}\right)-nN\\
\end{split}
\end{equation}
Then by applying the second part of  Theorem \ref{MAINA},  one obtains
\begin{equation}\label{eq:lem1eq2}
\begin{split}
&\grk\left(\begin{array}{cc}
zI-\mathbf A & -\mathbf B \\
\mathbf C & D
\end{array}\right)
=\\&N(n-1)+\grk \left(\begin{array}{cc}
q(z)I_N-p(z)L & -p(z)R\\
S& D \\
\end{array}\right)=\\
&N(n-1)+\grk \left(\begin{array}{cc}
h(z)I_N-L & -R\\
S& D \\
\end{array}\right)=\\
&N(n-1)+\grk \left(\begin{array}{cc}
\eta I_N-L & -R\\
S& D \\
\end{array}\right)=\\
&Nn+\grk\; \phi(z).
\end{split}
\end{equation}
By substituting the last equality of \eqref{eq:lem1eq2} into \eqref{eq:lem1eq1}, the result follows.

\hfill $\square \ $

\begin{theorem}\label{MAIN-B1}
Assume that  $(A_i,b_i,c_i)$ are SISO systems with identical transfer
function $g(z)=c_i(zI-A_i)^{-1}b_i$.
Then $(\mathbf{A},\mathbf{B},\mathbf{C},D)$  has
a zero at infinity if and only if $(L,R,S,D)$ has a
zero at infinity.
\end{theorem}

\noindent \textbf{Proof.}
By Lemma \ref{lem:normalrank}, the network transfer function matrix $\Gamma(z)$ and the interconnection transfer matrix $\phi(z)$ have the same normal rank. Using the conclusion of Theorem \ref{MAINA} (part \ref{partx}), the result follows.
\hfill $\square \ $

Theorem \ref{MAIN-B1}   shows that the infinite zero structure of
a homogeneous network depends only upon the interconnection parameters
and not on the specific details of the node transfer function.
This is in contrast to the finite zero structure, as is shown by the
following result.

\begin{theorem}\label{MAIN-B}
Assume that  $(A_i, b_i, c_i)$ are
SISO systems with identical transfer function $g(z)=c_i(zI-A_i)^{-1}b_i$. Let $p(z)/q(z)$ be a coprime polynomial factorization of $g(z)$ and define $h(z)=g(z)^{-1}$.  Let $(L,R,S,D)$ denote any constant interconnection matrices of the proper dimensions. \begin{enumerate}

\item
$(\mathbf{A},\mathbf{B},\mathbf{C},D)$  has a finite zero at $z_0\in
\mathbb{C}$ with $p(z_0)\neq 0$ if and only if  $h(z_0)\in
\mathbb{C}$ is a finite zero of
$(L,R,S,D)$.
\item $(\mathbf{A},\mathbf{B},\mathbf{C},D)$  has a finite zero at $z_0\in
\mathbb{C}$ with $p(z_0)= 0$ if and only if $(L,R,S,D)$ has a zero at infinity.
\end{enumerate}
\end{theorem}

\noindent \textbf{Proof.}
We first prove the first part of the theorem.  By Lemma \ref{lem:normalrank} and Theorem \ref{MAINA}, $z_0\in
\mathbb{C}$ is a zero of $(\mathbf{A},\mathbf{B},\mathbf{C},D)$ if and
only if
 \begin{equation}\label{eq:sisorank}
\rank \left(\begin{array}{cc}
q(z_0)I_N -p(z_0)L & -p(z_0)R\\
S & D \\
\end{array}\right)<N+\grk\;\phi (z).
\end{equation}
For $p(z_0)\neq 0$ this is equivalent to
\begin{equation*}
\rank \left(\begin{array}{cc}
h(z_0)I_N -L & -R\\
S & D \\
\end{array}\right)<N+\grk\;\phi (z),
\end{equation*}
i.e.  $h(z_0)$ being a finite zero of $(L,R,S,D)$.  For the second
part note that $z_0 \in \mathbb C$ is a zero of $(\mathbf{A},\mathbf{B},\mathbf{C},D)$ if and only if  inequality \eqref{eq:sisorank} holds.   If $p(z_0)=0$, then by
coprimeness of $p(z)$ and $q(z)$ we have $q(z_0)\neq 0$ and  therefore
\eqref{eq:sisorank} is equivalent to
\begin{equation*}
N+ \rank D=\rank \left(\begin{array}{cc}
q(z_0)I_N   & 0\\
S & D \\
\end{array}\right)<N+\grk\;\phi (z).
\end{equation*}
This is equivalent to $\rank\;D<\grk\;\phi (z)$.
Thus a zero of the node transfer function $g(z)$ is a zero of
$(\mathbf{A},\mathbf{B},\mathbf{C},D)$ if and only if $(L,R,S,D)$ has
a zero at infinity. This completes the proof.
\hfill $\square \ $

Now assume that $D$ has full-column rank or full-row rank. Then the
homogeneous network realization $(\mathbf{A},\mathbf{B},\mathbf{C},D)$
has no zeros at infinity. Thus in this case the finite zeros of
$(\mathbf{A},\mathbf{B},\mathbf{C},D)$ are exactly the preimages of the finite
zeros of $(L,R,S,D)$ under the rational function $ h(z)$.
We conclude with a result that is useful for the design of networks
with prescribed zero properties. The result below bears a certain
similarity with a result by Fax and Murray \cite{fax2004}.
As shown by them, a formation of $N$ identical vehicles can be analyzed for stability by analyzing a single vehicle with the same dynamics modified   by only a scalar, which assumes values equal to the eigenvalues of the interconnection matrix. Such a result is to do with poles, linking those of the individual agent  and the overall system via the eigenvalues (which are pole-like) of the interconnection matrix.  Our result is to do with the zeros, but  still links those of the individual agent, those of the interconnection matrix (suitably interpreted) and those of the whole system.

With the help of the preceding results, we can now study two other
important properties of  networks, namely, losslessness and
passivity. It is well known, see  e.g. \cite{vaidyanathan1989role}
(Section II. B), that if all agent  transfer function matrices and the
system defined by the quadruple $(L,R,S,D)$ are lossless, then the
system \eqref{eq:system} is lossless. We now provide an
improvement of this result for the case of SISO agents.\\

Recall  that a strictly proper real rational transfer
function $g(z)$ is called lossless \cite{vaid93} if all poles of $g(z)$ are in the open
unit disc and $|g(z)|=1$ holds for all $|z|=1$. A key property used below is that $|g(z)|>1$ if $|z|<1$ and $|g(z)|<1$ if $|z|>1$.

\begin{theorem}\label{cor:homog-network} Assume that $D$ has full-column   rank or full-row rank. Then
\begin{enumerate}
\item The homogeneous network $(\mathbf{A},\mathbf{B},\mathbf{C},D)$
  has no zeros at infinity. A complex number $z_0$ is a finite zero of
  $(\mathbf{A},\mathbf{B},\mathbf{C},D)$ if and only if $h(z_0)\neq\infty$ is a
  finite zero of $(L,R,S,D)$.
\item Assume that the agent transfer function $ g(z)$ is \textbf{lossless}. Then $(\mathbf{A},\mathbf{B},\mathbf{C},D)$ is a minimum phase
  network, i.e. all of its zeros have absolute value $<1$, if and only
  if $(L,R,S,D)$ is minimum phase.
\end{enumerate}
\end{theorem}

\noindent \textbf{Proof.}
The first claim is an immediate consequence of Theorem
\ref{MAIN-B}. If $g(z)$ is lossless then  $|g(z)|<1$ holds if and and
only if $|z|> 1$. Thus
$h(z)=1/g(z)$  maps the complement
of the open unit disc onto itself.  Thus $|z|\geq 1$ if and only
if $|h(z)|\geq 1$. Therefore $(L,R,S,D)$ has a
finite  zero
$\eta_0$ with $|\eta_0|\geq 1$ if and only if each $z$ with $h(z)=\eta_0$ satisfies
$|z|\geq 1$ and is a zero
of $(\mathbf{A},\mathbf{B},\mathbf{C},D)$. Note that for any finite $\eta_0$, there is necessarily a $z$  satisfying $h(z)=\eta_0$, since this is a polynomial equation for $z$.  This proves the result.
\hfill $\square \ $

We now extend the second part of the above corollary for  the choice of $\textbf{passive}$ transfer functions \cite{vaid93}.  Let us  recall that $g(z)$ is  passive  if and only if
\begin{enumerate}
\item
all poles of $g(z)$ are in $|z|\leq 1$
\item
$|g(z)|\leq1$  $\;\forall\;$ $|z|=1$.
\end{enumerate}
This implies
\begin{enumerate}
\item
$|g(z)|<1$   $\;\forall\;|z|>1$
\item
 If $|g(z)|>1$, then $|z|<1$.
\end{enumerate}

\begin{corollary}
Assume that $D$ has full-column rank or full-row rank and  $g(z)$ is \textbf{passive}. Then  $(\mathbf{A},\mathbf{B},\mathbf{C},D)$ is a minimum phase network, i.e. all of its zeros have absolute value $<1$, if $(L,R,S,D)$ is minimum phase.
\end{corollary}

\noindent \textbf{Proof.}
Suppose $|z_0|$ is a finite zero of $\{A,B,C,D\}$. Then $h(z_0)$ is a
finite zero of $(L,S,R,D)$, i.e.,  $1/g(z_0)$  is a finite zero of
$(L,S,R,D)$. By the minimum phase assumption, $|1/g(z_0)|<1$ or
$|g(z_0)|>1$. Passivity of $g(z_0)$ thus implies $|z_0|<1$.
\hfill $\square \ $
\bigskip

\subsection{Design of Networks}
An important issue is the construction of network topologies so that
the resulting  network is zero-free, i.e. it does not have
any finite zeros (but still may have a zero at infinity). We derive a
simple sufficient condition for homogeneous networks. By Corollary
\ref{cor:homog-network}, the homogeneous network
$(\mathbf{A},\mathbf{B},\mathbf{C},D)$ is zero-free if and only if $(L,R,S,D)$
is zero-free. For simplicity, we assume that there is a single
external input and a single external output associated with the
network, i.e. $m=p=1$. Moreover, we assume $D=0$. Thus the
interconnection transfer function $\phi(z)=S(zI_N-L)^{-1}R$ is
scalar strictly proper rational.
The next result characterizes which outputs of the SISO interconnected
system lead to a network without finite zeros, for given state and input
interconnection matrices.  

\begin{theorem}[SISO Design Condition]\label{DESIGN}
Assume that  $(A_i,b_i,c_i)$ are
identical minimal SISO systems with identical transfer function. Let $(L,R)$ be reachable with $L\in \mathbb{R}^{N\times N},
R\in \mathbb{R}^{N}$. Then a  network output
$S\in\mathbb{R}^{1\times N}$ defines a minimal network realization
 $(\mathbf{A},\mathbf{B},\mathbf{C},0)$ without finite zeros if and only if
 $S(zI_N-L)^{-1}R$ has relative degree $N$. 
\end{theorem}

\noindent \textbf{Proof.}
By Corollary \ref{cor:homog-network}, the homogeneous network
$(\mathbf{A},\mathbf{B},\mathbf{C},0)$  has no finite zeros
if and only if this holds for $(L,R,S,0)$. In the SISO case this
is equivalent to the transfer function $S(zI_N-L)^{-1}R$ having no
zeros. By \cite{fuhe2013}, $(\mathbf{A},\mathbf{B},\mathbf{C},0)$ is minimal if and only
if $(L,R,S)$ is minimal. In either case, $S(zI_N-L)^{-1}R$ has
McMillan degree $N$ and has no zeros if and only if the relative
degree of $S(zI_N-L)^{-1}R$ is equal to $N$.
\hfill $\square \ $

\begin{figure}[!t]
\begin{center}
    \includegraphics[width=5cm,height=2cm]{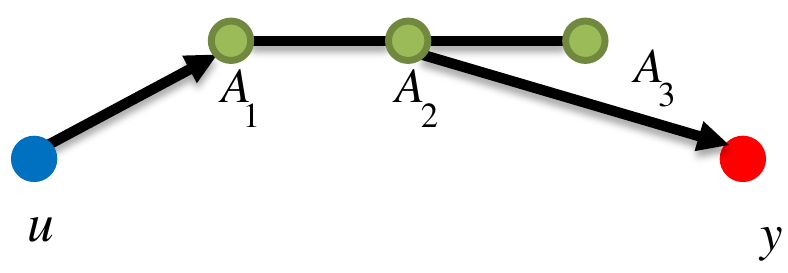}
    \caption{\emph{A homogenous network consisting of three SISO agents. The agents, the external input and measurement are depicted  by green, blue and red circles, accordingly. The whole network has two zeros at -1 and 1 when all weights are set to unity.}} \label{fig:threeagentszero}
\end{center}
\end{figure}

\begin{figure}[!t]
\begin{center}
    \includegraphics[width=5cm,height=2cm]{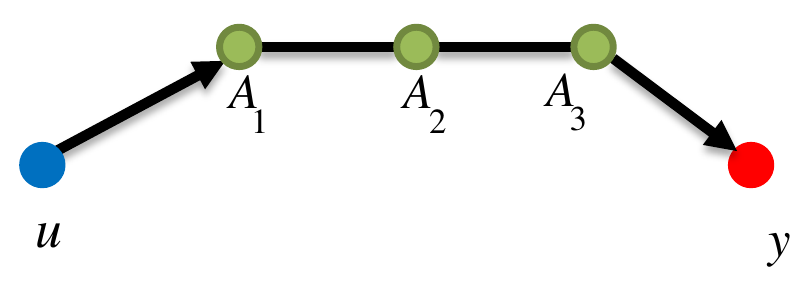}
    \caption{\emph{A homogenous network consisting of three SISO agents. The agents, the external input and measurement are depicted by green, blue and red circles, accordingly. The whole network is zero-free when all  weights are set to unity.}} \label{fig:threeagentsnozero}
\end{center}
\end{figure}

The above theorem characterizes when the SISO networked systems are
zero-free. We note that the condition is equivalent to the
sytem-theoretic condition that the closed loop
system $(\mathbf{A},\mathbf{B},\mathbf{C},0)$ is feedback
irreducible; i.e. that $(\mathbf{A}+\mathbf{B}K, \mathbf{C})$ is observable
for all state feedback matrices $K$.

The next example illustrates  that  the zeros of  the system (\ref{eq:system}) may drastically change  by replacing and adding   a link.

\begin{example}
Consider the network depicted in Fig. \ref{fig:threeagentszero} where the nodes are  simply double integrators. Note that there exist   bidirectional links between the  agents.  By assuming a unit weight on each link, it is easy to verify that for such a network  the interconnection matrices are  $ L=\left(\begin{array}{ccc}
1 &  1 & 0\\
 1 & 0 &  1\\
0 &  1 & 1
\end{array}\right), R=\left(\begin{array}{c}
1\\
0\\
0
\end{array}\right)$ and $S=\left(\begin{array}{ccc}
0 & 1 & 0 \end{array}\right)$. Moreover,  the interconnection dynamics
has a single zero at $z=1$. Hence, by using   Theorem \ref{MAIN-B} it is easy to see that the whole network has two zeros at $1$ and $-1$.   One can also observe that by adding an extra link in Fig. \ref{fig:threeagentszero} from  agent $A_3$  to the measurement node, with the same set of interconnection matrices  as before except for $S$ which assumes random values in its nonzero entries, the whole network becomes \textit{zero-free}. The same result holds i.e.  the resultant network is zero-free, when the topology  is modified according to Fig. \ref{fig:threeagentsnozero}.

\end{example}

\subsection{Circulant Homogeneous Networks}
Homogeneous networks with special coupling structures appear in many
applications, such as cyclic pursuit \cite{MarshallBrouckeFrancis};
shortening flows in image processing \cite{Bruckstein} or the
discretization of partial differential
equations~\cite{brockett-willems}.  Here, we characterize the zeros for
interconnections that have a  circulant
structure.
A homogeneous network is called \textbf{circulant}
if the state-to-state coupling matrix $L$ is a circulant, i.e.
\begin{equation*}
\begin{split}
 L&=\rm{Circ} (c_0,...,c_{N-1}) \\&=\begin{pmatrix}
                       c_0 & c_1 & \cdots &c_{N-2}& c_{N-1} \\
                       c_{N-1} & c_0 &c_1& \cdots & c_{N-2} \\
                       \vdots&\ddots&\ddots& \ddots&\vdots\\
                        c_2 & \cdots & c_{N-1} & c_0& c_{1} \\
                       c_1 & c_2&\cdots &c_{N-1}& c_0
                      \end{pmatrix}.
\end{split}
\end{equation*}
The book   \cite{davis} provides algebraic background on the circulant matrices.
A  basic fact on circulant matrices is that they are simultaneously diagonalizable by the \textbf{Fourier matrix}
\begin{equation*}
\Phi=\frac{1}{\sqrt{N}}\begin{pmatrix} 1&1&1& \ldots & 1\\
                         1& \omega &\omega^2 &\ldots & \omega^{N-1} \\
                         1 & \omega^2 & \omega^4 & \ldots & \omega^{2N-2}\\
                         \vdots & & & & \\
                         1 & \omega^{N-1} & \omega^{2N-2} & \ldots & \omega^{(N-1)^2}
                         \end{pmatrix} \text{ ,}
\end{equation*}
where $\omega=e^{2 \pi j /N}$  denotes a primitive $N-$th root of
unity. Note, that $\Phi$ is both a unitary and a symmetric
matrix.
It is then easily seen that any circulant matrix
$L$ has the form
$
 L=\Phi \diag(p_L(1),p_L(\omega),\ldots,p_L(\omega^{N-1}))\Phi^*,
$
 where
$
p_L(z):=\sum_{k=0}^{N-1}c_kz^{k-1}.
$
As a consequence of the preceding analysis we obtain the following result.

\begin{theorem} Suppose that the system in (\ref{eq:system}) is a circulant homogeneous network.  Let $D$ be full rank and $M= \diag(p_L(1),\ldots,p_L(\omega^{N-1}))$ and $w_1, \ldots, w_N$
denote the complex roots of
\begin{equation*}
\rm{det} \left(\begin{array}{cc}
wI_N -M & -\Phi^*R\\
S\Phi & D \\
\end{array}\right)=0.
\end{equation*}
Then
\begin{equation*}
\bigcup_{k=1}^{N}\{z\in \mathbb{C}\;|\; q(z)-w_kp(z)=0\}
\end{equation*}
are the finite zeros of the homogeneous network $(\mathbf{A},\mathbf{B},\mathbf{C},D)$.
\end{theorem}

\noindent \textbf{Proof.}
By Theorem \ref{MAIN-B}, we  conclude that  the  system defined by $(\mathbf{A},\mathbf{B},\mathbf{C},D)$ has a finite nonzero zero  if and only if the following matrix pencil has less than full rank
\begin{equation}\label{eq:h(z)}
\left(\begin{array}{cc}
h(z)I_N -L & -R\\
S & D \\
\end{array}\right).
\end{equation}
Observe that the following equality holds
\begin{equation}
\begin{split}
&\left(\begin{array}{cc}
h(z)I_N -M & -\Phi^*R\\
S\Phi & D \\
\end{array}\right)=\\& \left(\begin{array}{cc}
\Phi^{*} & 0\\
0 & I \\
\end{array}\right)  \left(\begin{array}{cc}
h(z)I_N -L & -R\\
S & D \\
\end{array}\right) \left(\begin{array}{cc}
\Phi & 0\\
0 & I \\
\end{array}\right) .
\end{split}
\end{equation}
Note that multiplication of a matrix by non-singular matrices on the left and right respectively does not change the rank. This implies the result.
 \hfill $\square \ $

\section{Zeros of Blocked Networked Systems}\label{sec:blocking}

The technique of blocking or lifting  a signal is well-known in
systems and control \cite{chenB95} and signal processing
\cite{vaid93}. In systems theory,  this  method
has been
mostly  exploited to  transform linear discrete-time periodic
systems into
linear time-invariant systems in order to apply the well-developed  tools for  linear time-invariant systems; see
\cite{bittanti09} and the literature therein. Here, we show how this
technique can be applied to the networked  systems of the form
\begin{equation}\label{eq:eq3}
\begin{split}
x(t+1) &= {\mathbf{A}}x(t)+ {\mathbf{B}}u(t)\\
y(t) &= {\mathbf{C}}x(t)+Du(t),
\end{split}
\end{equation}
with matrices
\begin{equation*}
\mathbf{A} := A + B L C
\quad
\mathbf{B} := B R,
\quad
{\mathbf{C}} := SC.
\end{equation*}
and the  network transfer function
\begin{equation*}
\Gamma (z)= D+SC(z I-A-BLC)^{-1}BR.
\end{equation*}
Here $x(t) \in \mathbb{R}^{\overline{n}}$, $y(t) \in \mathbb{R}^{p}$ and $ u(t) \in
\mathbb{R}^{m}$ and $A=\diag (A_1,\ldots, A_N),B=\diag
(B_1,\ldots, B_N),C=\diag (C_1,\ldots, C_N)$ are block-diagonal.
Given an integer $T\geq 1$ as the block size,  we define for $t=0,T,2T,\ldots$
 \begin{equation}
\begin{split}
U(t)  &= \left(\begin{array}{cccc}
{u(t)}^{\top}& {u(t+1)}^{\top} & \ldots & {u(t+T-1)}^{\top}\end{array}\right)^{\top},\nonumber\\ Y(t)&=\left(\begin{array}{cccc}
{y(t)}^{\top}& {y(t+1)}^{\top} & \ldots & {y(t+T-1)}^{\top}\end{array}\right)^{\top}.
\end{split}
\end{equation}
The \textbf{blocked system}  then is defined as \cite{bittanti09}
\begin{equation}
\begin{split}
x(t+T)& =  {\mathbf{A}}_bx(t)+{\mathbf{B}}_b U(t)\\
Y(t) & =   {\mathbf{C}}_bx(t)+{\mathbf{D}}_b U(t),
\end{split}\label{eq:eq6}
\end{equation}
where
\begin{eqnarray}
{\mathbf{A}}_b &=& \mathbf{A}^{T},\quad {\mathbf{B}}_b=\left(
 \begin{array}{cccc} \mathbf{A}^{T-1}\mathbf{B} & \mathbf{A}^{T-2}\mathbf{B} & \ldots & \mathbf{B} \end{array}\right),\nonumber\\
 {\mathbf{C}}_{b}&=&\left( \begin{array}{cccc}\mathbf{C}^{\top}&
 \mathbf{A}^{\top} \mathbf{C}^{\top}& \ldots & \mathbf{A}^{(T-1)^{\top}}\mathbf{C}^{\top}\end{array}\right)^{\top},\nonumber\\
 {\mathbf{D}}_b&=&\left(\begin{array}{cccc} D & 0 & \ldots & 0\\
\mathbf{C}\mathbf{B}& D & \ldots & 0 \\
\vdots & \vdots & \ddots & \vdots \\
\mathbf{C}\mathbf{A}^{T-2}\mathbf{B}& \mathbf{C}\mathbf{A}^{T-3}\mathbf{B} & \ldots & D\end{array}\right).
\label{eq:eq10}
\end{eqnarray}
The transfer function
$
\Gamma_b(z^T)={\mathbf{D}}_b+{\mathbf{C}}_b(z^T I-{\mathbf{A}}_b)^{-1}{\mathbf{B}}_b$ of (\ref{eq:eq3}),
 see \cite{bittanti09}, \cite{Khargoneckar85}, has the circulant-like
 structure as

{\footnotesize
\begin{equation*}
\label{eq:blockedtransfer}
\begin{pmatrix}
 H_0(z)& H_{T-1}(z) & \ldots&H_{2}(z) & H_{1}(z) \\
zH_{1}(z) & H_0(z)&H_{T-1}(z)&\dots & H_{2}(z)   \\
\vdots & \ddots & \ddots & \ddots& \vdots\\
zH_{T-2}(z)&\dots&zH_{1}(z)&H_0(z)&H_{T-1}(z)\\
zH_{T-1}(z) & zH_{T-2}(z) & \dots&zH_{1}(z)& H_0(z)
\end{pmatrix}
\end{equation*} }
where $H_{0}(z)=D+{\mathbf{C}}(z I-{\mathbf{A}}^{T})^{-1}{\mathbf{A}}^{T-1}{\mathbf{B}}$ and
$H_{k}(z)=\mathbf{C}(z I-\mathbf{A}^{T})^{-1}\mathbf{A}^{k-1}\mathbf{B}, k=1,\dots,T-1$.
It is worthwhile mentioning  that the blocked transfer function has the
structure of a generalized circulant matrix.  The theory of generalized
circulant matrices is very similar to that of classical circulant
matrices; see \cite{davis}. Using such techniques we obtain the following result.

In order to deal with the zeros of the system \eqref{eq:eq6},  we  first  need to review the  following result  from \cite{Varga2003}, obtained by specializing Lemma 1 of \cite{Varga2003} to the time-invariant case.
\begin{lemma}\label{lemvarga} \cite{Varga2003}
Let $\widetilde{\mathbf A}_b=I_T\otimes \mathbf A$, $\widetilde{\mathbf B}_b=I_T\otimes \mathbf B $, $\widetilde{\mathbf C}_b=I_T \otimes \mathbf C$ and $\widetilde{ \mathbf D}_b=I_T \otimes \mathbf D$. Furthermore,  define $\mathbf E_\zeta \triangleq \left(
                                                                                                                                                                                                           \begin{array}{cccc}
                                                                                                                                                                                                             0 & 1 &  & 0 \\
                                                                                                                                                                                                             0 &  & \ddots &  \\
                                                                                                                                                                                                              \vdots &  & \ddots & 1 \\
                                                                                                                                                                                                             \zeta & 0 &  & 0 \\
                                                                                                                                                                                                           \end{array}
                                                                                                                                                                                                         \right)
 $, $\mathbf E_\zeta \in \mathbb C^{ T \times T}$ and
 $\widetilde{\mathbf E}_\zeta= \mathbf E_\zeta \otimes I_{\bar n}$ where
 $\otimes$ denotes the Kronecker product and $\zeta$ denotes a complex number. Then there exist invertible  matrices $T_l$ and $T_r$ and  matrices $X$ and $Y$   such that for all $\zeta \in \mathbb{C}$
\begin{equation}\label{eq:vargelemmaeq}
\begin{split}
& \left(\begin{array}{ccc}I_{\bar n(T-1)}& 0& 0\\ 0& \zeta I-\mathbf A_{b} & -\mathbf B_{b}\\
                             0& \mathbf C_{b}& \mathbf D_{b}
                             \end{array}\right)=\\&
\left(\begin{array}{cc} T_l & 0\\
                            X & I  \end{array}\right) \left(\begin{array}{cc} \widetilde{\mathbf E}_\zeta-\widetilde{\mathbf A}_{b} & -\widetilde{\mathbf B}_{b}\\
                             \widetilde{\mathbf C}_{b} &
                             \widetilde{\mathbf D}_{b}
                             \end{array}\right)
                             \left(\begin{array}{cc} T_r & Y\\
0 & I  \end{array}\right).
\end{split}
\end{equation}
\end{lemma}

Using this lemma we introduce the following result.

\begin{proposition}\label{PROPOSITION}
Let $\Phi$ denote the Fourier matrix of the  proper dimension and $\Gamma (z)=Q(z)^{-1}P(z)$ be a left
coprime factorization of the network transfer function.
Consider the system matrices
\begin{equation*}
\begin{split}
&\Sigma_b(z)=\begin{pmatrix}
zI_{\overline{n}}-{\mathbf{A}}_b & -{\mathbf{B}}_b\\
{\mathbf{C}}_b & {\mathbf{D}}_b
\end{pmatrix},\\  &\hat{\Sigma}_b(z)=\begin{pmatrix}
I_{ \overline{n}(T-1)} & 0\\
0 & \Sigma_b(z)
\end{pmatrix}.
\end{split}
\end{equation*}
There exist invertible  matrices
$L (z)$ and $R(z)$ that  are invertible for all nonzero complex numbers
$z\in\mathbb{C}$  such that
{\footnotesize
\begin{equation}
\label{eq:blockedtransferdiag1}
\begin{split}
&\hat{\Sigma}_b(z^T)=\\&{ \scriptsize L(z)\begin{pmatrix}
\begin{pmatrix}
zI_{ \overline{n}}-\mathbf{A} & -\mathbf{B}\\
\mathbf{C} &  D
\end{pmatrix} & & 0 \\
&   \ddots  & \\
0 &  & \begin{pmatrix}
\omega^{T-1} zI_{\overline{n}}- \mathbf{A} & -\mathbf{B}\\
\mathbf{C} & D
\end{pmatrix}
\end{pmatrix}R(z)}
\end{split}
\end{equation}
}
%
\end{proposition}

\noindent \textbf{Proof.}
First, observe that the following equality holds
\begin{equation}\label{eq:circul}
\mathbf E_1=\Phi\left(\begin{array}{llll}
1&   &  &  \\
 & \omega &   &  \\
 &  & \ddots &  \\
 &  &  & \omega^{T-1}
\end{array}\right)\Phi^*,
\end{equation}
where $\Phi$ is the Fourier matrix of the proper dimension. Furthermore, we have

\begin{equation}\label{eq:circul2}
\mathbf E_\zeta=z\Delta (z)\mathbf E_1\Delta(z)^{-1}.
\end{equation}
where $\Delta(z)=\left(\begin{array}{llll}
1 &  &   &  \\
 & z  &  &  \\
 &  & \ddots & \\
 &  &  &z^{T-1}
\end{array}\right)$.

Now  by using  \eqref{eq:circul2} and  \eqref{eq:circul}, one can easily verify that  the following equality holds
\begin{equation}
\begin{split}
\widetilde{\mathbf E}_\zeta=& (\Delta(z)\Phi) \otimes I_{\bar n} \left(\begin{array}{llll}
zI_{\bar n}&   &  &  \\
 & \omega zI_{\bar n} &   &  \\
 &  & \ddots &  \\
 &  &  & \omega^{T-1}zI_{\bar n}
\end{array}\right)\\&(\Delta(z)\Phi)^{-1} \otimes I_{\bar n}.
\end{split}
\end{equation}

Therefore, for any $\zeta\ne0$, $z ^T=\zeta$, we have $\bar T(z) \triangleq \Delta(z) \Phi \otimes I_{\bar n}$, $\bar R(z) \triangleq \Delta(z)\Phi \otimes I_m$ and $\bar L(z)\triangleq \Delta(z)\Phi \otimes I_p$. Hence,
\begin{equation}
\begin{split}\label{eq:nice}
&\left(\begin{array}{cc} \widetilde{\mathbf E}_\zeta-\widetilde{\mathbf A}_{b} & -\widetilde{\mathbf B}_{b}\\
                             \widetilde{\mathbf C}_{b} & \widetilde{\mathbf D}_{b}  \end{array}\right)=
\left(\begin{array}{r|lc} \bar T(z) &0 \\ \hline 0 &\bar L(z) \end{array}\right)\\&\left(\begin{array}{r|lc} \left(\begin{array}{ccc}z I-A & &0\\
& \ddots& \\
0&&\omega^{T-1} z I-A \end{array}\right)  & \left(\begin{array}{ccc}B & &0\\
& \ddots& \\
0&&B \end{array}\right)\\\\ \hline\\ \left(\begin{array}{ccc}C & &0\\
& \ddots& \\
0&&C\end{array}\right) &\left(\begin{array}{ccc}D & &0\\
& \ddots& \\
0&&D\end{array}\right)\end{array}\right)\\&\left(\begin{array}{r|lc} \bar T^{-1}(z) &0 \\ \hline 0 &\bar R^{-1}(z) \end{array}\right).
\end{split}
\end{equation}

Now by substituting (\ref{eq:nice}) into the equation \eqref{eq:vargelemmaeq} and performing the required rows and  columns reordering,  the conclusion of the proposition becomes immediate. 

\hfill $\square \ $

The preceding results imply the following  characterization  of
the finite zeros for the interconnected systems.  Thus consider the
interconnected system  $(\mathbf{A},\mathbf{B},\mathbf{C},D)$ defined
in (\ref{eq:system}). Let
$(\mathbf{A_b},\mathbf{B_b},\mathbf{C_b},D_b)$ denote the associated
blocked system, defined as in \eqref{eq:eq6} and \eqref{eq:eq10}.
\begin{theorem}
 A complex number $\zeta_0\neq 0$ is a finite zero of the blocked
network
$(\mathbf{A_b},\mathbf{B_b},\mathbf{C_b},D_b)$ if and only if there
exists $z_0\in \mathbb{C}$ with $z_0^T=\zeta_0$ such that $z_0$ is a finite zero of
$(\mathbf{A},\mathbf{B},\mathbf{C},D)$.
\end{theorem}
 \noindent \textbf{Proof.}
  \textbf{Necessity.} Suppose that $\zeta_0=z_0^T$ is a zero of the system matrix $\Sigma_b(\zeta)$, then by recalling  the result  of Proposition \ref{PROPOSITION}, one can easily  see  that one or  more of diagonal blocks in \eqref{eq:blockedtransferdiag1} should have rank below their  normal rank i.e. there exist at least one $T$-th root of $\zeta_0$ which is a zero  of the unblocked system. \\
  \textbf{Sufficiency.}  Suppose that $z_0$ is a zero of the unblocked system \eqref{eq:eq3}. Then at least one of the diagonal blocks in \eqref{eq:blockedtransferdiag1} loses rank below its normal rank. Now, again  by using   \eqref{eq:blockedtransferdiag1}, one can conclude that $\zeta_0=z_0^T$ is a zero of $\hat{\Sigma}_b(z)$. The latter implies that $\zeta_0$ must be a zero of the system $\eqref{eq:eq6}$.

\hfill $\square \ $

The above theorem only treats the finite nonzero zeros. To treat the
other cases i.e. zeros at the origin and infinity, we recall the following
result from \cite{MOHSEN-SCLpaper}.

\begin{proposition}\label{thm:zeroatinfinity}
Consider the  unblocked networked  system \eqref{eq:eq3} with  transfer
function  $\Gamma(z)$ and the  blocked networked  system  (\ref{eq:eq6})
with transfer function $\Gamma_b(\zeta)$. Suppose that  the quadruple
$(\mathbf A,\mathbf B,\mathbf C,D)$ is minimal. Then
\begin{enumerate}
\item The system \eqref{eq:eq3} has a zero at  $z=\infty$  if  and only if the system (\ref{eq:eq6})  has a zero at $\zeta=\infty$.
\item The system \eqref{eq:eq3} has a zero at the origin if and only if the   the system (\ref{eq:eq6})  has a zero at  the origin.
\end{enumerate}
\end{proposition}

This implies the next characterization of the zeros for the systems \eqref{eq:eq3} and (\ref{eq:eq6}).

\begin{theorem}
Let $D$ be full rank and $(\mathbf{A},\mathbf{B},\mathbf{C},D)$
a homogeneous network with SISO agents. Then the blocked network
$(\mathbf{A_b},\mathbf{B_b},\mathbf{C_b},D_b)$
has no zeros at infinity.  The finite zeros of
$(\mathbf{A_b},\mathbf{B_b},\mathbf{C_b},D_b)$ are exactly all
$\zeta=z^T$ such that $ h(\omega^{k}z)$ is a finite zero of $(L,R,S,D)$ for some
 $0 \leq k \leq T-1$.
\end{theorem}

 \noindent \textbf{Proof.}
The proof readily follows from Proposition \ref{thm:zeroatinfinity}
 and the first part of Theorem \ref{cor:homog-network}.
 \hfill $\square \ $

\section{Conclusions}\label{sec:conclusion}
In this paper, we explored the zeros of  networks of linear
systems. It was assumed that the interaction topology is time-invariant.  The zeros  were  characterized for both homogeneous and
heterogeneous networks. In particular, it was  shown that for homogeneous
networks with  full rank direct feedthrough
matrix,  the finite zeros of the whole network  are
exactly the preimages of interconnection dynamics zeros  under the  inverse of an agent transfer
function. We then discussed the condition under which the networked systems  have  no finite nonzero zeros. Then
\textit{ generalized circulant matrices} were used for a concise
analysis of the  finite nonzero zeros of blocked networked  systems. Moreover, we recalled some results about their zeros at infinity and at the origin.  It was shown that the  networked systems have zeros at the origin  (infinity) if and only if their associated blocked systems have zeros at the origin (infinity). As a part of our future work  we will
address open problems such as the consideration of periodically varying network topologies and MIMO dynamics for each agents. Furthermore, as explained in the  illustrative example given in the current paper,  adding and removing links can dramatically change the  zero  structure. Thus, another interesting  research direction involves exploring how links in the  networked systems  can be  systematically  designed  such that the resultant networked systems   attain   a particular zero dynamics.

\bibliographystyle{plain}
\bibliography{acc2012bib2}           

\begin{thebibliography}{10}

\bibitem{amin:09}
S.~Amin, A.~A. C{\'a}rdenas, and S.~Sastry.
\newblock Safe and secure networked control systems under denial-of-service
  attacks.
\newblock In {\em Hybrid Systems: Computation and Control}, pages 31--45.
  Springer, 2009.

\bibitem{bittanti09}
S.~Bittanti and P.~Colaneri.
\newblock {\em Periodic Systems Filtering and Control}.
\newblock Communications and Control Engineering. Springer-Verlag, 2009.

\bibitem{Bolzern86}
P.~Bolzern, P.~Colaneri, and R.~Scattolini.
\newblock Zeros of discrete-time linear periodic systems.
\newblock {\em IEEE Transactions on Automatic Control}, 31(11):1057--1058,
  1986.

\bibitem{brockett-willems}
R.~W. Brockett and J.~L. Willems.
\newblock Discretized partial differential equations: Examples of control
  systems defined on modules.
\newblock {\em Automatica}, 10(5):507--515, 1974.

\bibitem{Bruckstein}
M.~A. Bruckstein, G.~Sapiro, and D.~Shaked.
\newblock Evolutions of planar polygons.
\newblock {\em International Journal of Pattern Recognition and Artificial
  Intelligence}, 9(6):991--1014, 1995.

\bibitem{cardenas2011attacks}
A.~A. C{\'a}rdenas, S.~Amin, Z.~Lin, Y.~Huang, C.~Huang, and S.~Sastry.
\newblock Attacks against process control systems: risk assessment, detection,
  and response.
\newblock In {\em ACM Symposium on Information, Computer and Communications
  Security}, pages 355--366, 2011.

\bibitem{chenB95}
T.~Chen and B.~A. Francis.
\newblock {\em Optimal Sampled-Data Control Systems}.
\newblock Springer-Verlag New York, Inc., Secaucus, NJ, USA, 1995.

\bibitem{chen2010}
W.~Chen, B.~D.~O. Anderson, M.~Deistler, and A.~Filler.
\newblock Properties of blocked linear systems.
\newblock {\em Automatica}, 48(10):2520--2525, 2012.

\bibitem{davis}
P.~J. Davis.
\newblock {\em Circulant matrices}.
\newblock John Wiley and Sons. New York, 1979.

\bibitem{falliere2011w32}
N.~Falliere, L.~O. Murchu, and E.~Chien.
\newblock W32. stuxnet dossier.
\newblock {\em White paper, Symantec Corp., Security Response}, 2011.

\bibitem{fax2004}
J.~A. Fax and R.~M. Murray.
\newblock Information flow and cooperative control of vehicle formations.
\newblock {\em IEEE Transactions on Automatic Control}, 49(9):1465--1476, 2004.

\bibitem{Fuhrmann1977}
P.~A. Fuhrmann.
\newblock On strict system equivalence and similarity†.
\newblock {\em International Journal of Control}, 25(1):5--10, 1977.

\bibitem{fuhe2013}
P.~A. Fuhrmann and U.~R. Helmke.
\newblock Observability and strict equivalence of networked linear systems.
\newblock {\em Mathematics of Control, Signals, and Systems}, 25(4):437--471,
  2013.

\bibitem{gorman2009electricity}
S.~Gorman.
\newblock Electricity grid in us penetrated by spies.
\newblock {\em The Wall Street Journal}, 8, 2009.

\bibitem{Grasselli88}
O.~M. Grasselli and S.~Longhi.
\newblock Zeros and poles of linear periodic multivariable discrete-time
  systems.
\newblock {\em Circuits, Systems, and Signal Processing}, 7:361--380, 1988.

\bibitem{gupta2010optimal}
A.~Gupta, C.~Langbort, and T.~Basar.
\newblock Optimal control in the presence of an intelligent jammer with limited
  actions.
\newblock In {\em IEEE Conference on Decision and Control}, pages 1096--1101,
  2010.

\bibitem{kailath}
T.~Kailath.
\newblock {\em Linear Systems}.
\newblock Prentice-Hall, New Jersey, 1980.

\bibitem{Khargoneckar85}
P.~Khargonekar, K.~Poolla, and A.~Tannenbaum.
\newblock Robust control of linear time-invariant plants using periodic
  compensation.
\newblock {\em IEEE Transactions on Automatic Control}, 30(11):1088--1096,
  1985.

\bibitem{MarshallBrouckeFrancis}
J.~A. Marshall, M.~E. Broucke, and B.~A. Francis.
\newblock Formations of vehicles in cyclic pursuit.
\newblock {\em IEEE Transactions on Automatic Control}, 49(11):1963--1974,
  2004.

\bibitem{sinapoli2012}
Y.~Mo, T.~H.~H. Kim, K.~Brancik, D.~Dickinson, L.~Heejo, A.~Perrig, and
  B.~Sinopoli.
\newblock Cyber-physical security of a smart grid infrastructure.
\newblock {\em Proceedings of the IEEE}, 100(1):195--209, 2012.

\bibitem{Olfati2007}
R.~Olfati-Saber, J.~A. Fax, and R.~M. Murray.
\newblock Consensus and cooperation in networked multi-agent systems.
\newblock {\em Proceedings of the IEEE}, 95(1):215--233, 2007.

\bibitem{olfati2002distributed}
R.~Olfati-Saber and R.~M. Murray.
\newblock Distributed cooperative control of multiple vehicle formations using
  structural potential functions.
\newblock In {\em IFAC World Congress}, pages 346--352, 2002.

\bibitem{Ren2007}
W.~Ren, R.~W. Beard, and E.~M. Atkins.
\newblock Information consensus in multivehicle cooperative control.
\newblock {\em IEEE Control Systems Magazine}, 27(2):71--82, april 2007.

\bibitem{rid2012cyber}
T.~Rid.
\newblock Cyber war will not take place.
\newblock {\em Journal of Strategic Studies}, 35(1):5--32, 2012.

\bibitem{Rosenbrock1970}
H.~H. Rosenbrock.
\newblock {\em State-Space and Multivariable Theory}.
\newblock Nelson, London, UK, 1970.

\bibitem{sinopoli2003distributed}
B.~Sinopoli, C.~Sharp, L.~Schenato, S.~Schaffert, and S.~Sastry.
\newblock Distributed control applications within sensor networks.
\newblock {\em Proceedings of the IEEE}, 91(8):1235--1246, 2003.

\bibitem{govinndarasu}
S.~Sridhar, A.~Hahn, and M.~Govindarasu.
\newblock Cyber-physical system security for the electric power grid.
\newblock {\em Proceedings of the IEEE}, 100(1):210--224, 2012.

\bibitem{tanner2003stable}
H.~G. Tanner, A.~Jadbabaie, and G.~J. Pappas.
\newblock Stable flocking of mobile agents part i: dynamic topology.
\newblock In {\em IEEE Conference on Decision and Control}, pages 2016--2021,
  2003.

\bibitem{Teixeira202}
A.~Teixeira, I.~Shames, H.~Sandberg, and K.H. Johansson.
\newblock Revealing stealthy attacks in control systems.
\newblock In {\em Allerton Conference on Communication, Control, and
  Computing}, pages 1806--1813, 2012.

\bibitem{vaid93}
P.~P. Vaidyanathan.
\newblock {\em {Multirate Systems and Filter Banks}}.
\newblock Prentice-Hall, Inc., Upper Saddle River, NJ, USA, 1993.

\bibitem{vaidyanathan1989role}
P.~P. Vaidyanathan and Z.~Doganata.
\newblock The role of lossless systems in modern digital signal processing: A
  tutorial.
\newblock {\em IEEE Transactions on Education}, 32(3):181--197, 1989.

\bibitem{Varga2003}
A.~Varga and P.~{ Van Dooren}.
\newblock Computing the zeros of periodic descriptor systems.
\newblock {\em Systems and Control Letters}, 50(5):371--381, 2003.

\bibitem{MOHSEN-SCLpaper}
M.~Zamani, B.~D.~O. Anderson, U.~R. Helmke, and W.~Chen.
\newblock On the zeros of blocked time-invariant systems.
\newblock {\em Systems and Control Letters}, 62(7):597--603, 2013.

\bibitem{zamani2011}
M.~Zamani, W.~Chen, B.~D.~O. Anderson, M.~Deistler, and A.~Filler.
\newblock On the zeros of blocked linear systems with single and mixed
  frequency data.
\newblock In {\em IEEE Conference on Desicion and Control}, pages 4312--4317,
  2011.

\bibitem{zamanihlemke2013}
M.~Zamani, U.~R. Helmke, and B.~D.~O. Anderson.
\newblock zeros of networks of linear multi-agent systems: time-invariant
  interconnections.
\newblock In {\em IEEE Conference on Decision and Control}, pages 5939--5944,
  2013.

\end{thebibliography}

\end{document}